\documentclass[ superscriptaddress]{revtex4-2}
\usepackage{epstopdf}
\usepackage{graphicx}
\usepackage{dcolumn}
\usepackage{amsmath}
\usepackage{amsfonts}
\usepackage{epsfig}
\usepackage{subcaption}
\graphicspath{ {./image/} }
\usepackage{amssymb}
 
 \usepackage{hyperref}
  
\begin{document}
 
\title{More Exact Thermodynamics of Nonlinear Charged AdS Black Holes in 4D
Critical Gravity}
\author{Prosenjit Paul}\email{prosenjitpaul629@gmail.com}
\affiliation{Indian Institute of Engineering Science and Technology (IIEST), Shibpur-711103, WB, India}
\author{Sudhaker Upadhyay\footnote{Corresponding author}\footnote{Visiting Associate, IUCAA Pune, Maharashtra-411007, India}}\email{sudhakerupadhyay@gmail.com}
\affiliation{Department of Physics, K.L.S. College, Magadh University, Nawada, Bihar 805110, India}
\affiliation{School of Physics, Damghan University, Damghan, 3671641167, Iran}
\author{Yerlan Myrzakulov}\email{ymyrzakulov@gmail.com}
\affiliation{Department of General \& Theoretical Physics,
LN Gumilyov Eurasian National University, Astana, 010008, Kazakhstan}
\author{Dharm Veer Singh\email{Visiting Associate, IUCAA Pune, Maharashtra-411007, India} }\email{veerdsingh@gmail.com}
\affiliation{Department of Physics, Institute of Applied Sciences and Humanities,
GLA University, Mathura 281406, Uttar Pradesh, India.}
\author{Kairat Myrzakulov}\email{krmyrzakulov@gmail.com}
\affiliation{Department of General \& Theoretical Physics,
LN Gumilyov Eurasian National University, Astana, 010008, Kazakhstan}

\begin{abstract}
    In this paper, we investigate nonlinearly charged AdS black holes in four-dimensional critical gravity and study more exact black hole thermodynamics under the effect of small statistical fluctuations. We compute the correction to the thermodynamics of nonlinearly charged AdS black hole up to the leading order. We discuss the stability of  black holes under the circumstances of fluctuation and find that fluctuation causes instability in the black holes. Moreover, both the isothermal  and adiabatic
compressibilities are also derived. 
Finally, we estimate the role of small fluctuations on the equation of states and  study the $P-v$ diagram of nonlinearly charged AdS black hole.
\end{abstract}

\maketitle

\section{Overview and motivation}

The history of black hole thermodynamics is quite long. In 1972, Bekenstein conjectured that black
holes possess an entropy \cite{bekenstein2020black}. Later, in 1973, a relationship between black hole entropy and horizon area is established by Bekenstein \cite{bekenstein1973black}. There it was found that the   black hole entropy is proportional to the area of the event
horizon. In 1973, Bardeen, Carter, and Hawking proposed the four-laws of black hole mechanics \cite{bardeen1973four} following the analogy with the four-laws of thermodynamics. In 1974, Hawking \cite{hawking1974black,hawking1975particle} proposed that black holes emit thermal radiation and the temperature of the radiation is inversely proportional to black hole mass. Hawking and Page found a black hole solution in asymptotically AdS space \cite{hawking1983thermodynamics} which possesses the thermodynamics properties like entropy, temperature, etc. 

Black hole physics has been a fascinating subject of study for several decades and their thermodynamics has been an active area of research. Jacob Bekenstein proposed the so-called ``entropy-area" law, which suggested that the entropy of a black hole is proportional to the area of its event horizon. Subsequent research in the 1990s and 2000s focused on understanding the quantum corrections to black hole entropy that arise due to thermal fluctuations and other quantum effects. The Cardy formula 
\cite{Kaul:2000kf,Carlip:2000nv}, introduced by John Cardy in 1986, stands out as one of the most significant breakthroughs in this domain. The Cardy formula provides a powerful tool for studying the connection between black hole thermodynamics and conformal field theory and has been used extensively to study black hole entropy. Another important development was the discovery of the AdS/CFT correspondence in 1997 by   Maldacena. This correspondence provides a way to understand the behavior of quantum systems in curved space-time, such as that near a black hole, in terms of a dual quantum field theory living on the boundary of space-time. This has led to important insights into the nature of black hole entropy and the holographic principle, which suggests that the properties of a system can be understood in terms of its boundary degrees of freedom.   Other interesting  black hole solutions are also studied \cite{an1,an2,an3,an4}.

In 2002, Das computed \cite{das2002general} logarithmic corrections to the entropy. In recent years, there has been growing interest in studying the  logarithmic corrections to the entropy of black holes due to small statistical fluctuations around black hole equilibrium. Assuming that a black hole behaves as a thermodynamic system and this system should follow the equilibrium with thermal radiation. However, the logarithmic corrections to thermodynamic entropy arise for all thermodynamic systems when small statistical fluctuations around equilibrium are taken into account \cite{das2002general}. A nontrivial multiplicative factor to the expression for the density of states arises due to the small statistical fluctuations and the logarithm of these multiplicative factors leads to the corrections to the entropy. Thus, Bekenstein-Hawking entropy can be modified by logarithmic corrections that result from thermal fluctuations of the black hole around its state of equilibrium. These logarithm corrections to the entropy of the black hole are universal and apply to all kinds of black hole spacetime irrespective of whether they arise in Einstein's gravity or any higher-order theories of gravity. The inclusion of logarithmic corrections to the Bekenstein-Hawking entropy can be understood as the thermal fluctuations experienced by the black hole as it deviates from its stable state.

Recent studies have demonstrated that thermal fluctuations play a crucial role in understanding the behaviour of charged anti-de Sitter black holes \cite{Pourhassan:2015cga}, leading to corrections in their thermodynamic properties. In fact, a thorough analysis of black holes has revealed that the quantum approach to their thermodynamics at small scales is essential, resulting in a variety of corrections to thermodynamic quantities. Investigations into the effects of such corrections have been conducted for a range of black holes, including the Godel black hole \cite{Pourdarvish:2013gfa}, quasitopological black holes \cite{upadhyay2017quantum} and the Schwarzschild–Beltrami–de Sitter black hole \cite{Pourhassan:2017rie}, charged rotating black holes in AdS space \cite{upadhyay2018leading},  Horava-Lifshitz black holes \cite{pourhassan2018quantum,pourhassan}, charged black holes in gravity rainbow \cite{upadhyay2018thermal}, black holes in $f(R)$ gravity \cite{upadhyay2021perturbed}, rotating and charged BTZ black hole \cite{upadhyay2022modified} and 
  Schwarzschild black hole
immersed in holographic quintessence \cite{sara} etc. The pioneering research of Frolov \cite{Frolov:1996hd} has provided significant insights into the quantum corrections to black hole thermodynamics. The logarithmic corrections to the entropy of black holes have important consequences. One of the most significant is that they violate the area law of black hole entropy. The area law states that the entropy of a black hole is proportional to the area of its event horizon. However, the logarithmic corrections introduce additional terms in the entropy that are not proportional to the area of the event horizon. 

The theory of nonlinear electrodynamics was first proposed \cite{Born:1934ji,Born:1934gh,Born:1935ap,Infeld:1936wzo,Infeld:1937frv} by Born and Infeld to remove the singularity of electromagnetic fields due to point particle. After Born-Infeld electrodynamics, a new model of nonlinear electrodynamics was proposed by Plebanski using antisymmetric conjugate tensor $P^{\mu \nu}$ (known as Plebanski tensor) and a structure-function $\mathcal{H} = \mathcal{H}(P, Q)$, where $P$  and $Q$ are the invariants formed with the antisymmetric conjugate Plebanski tensor. The structure-function $\mathcal{H}(P,Q)$ is related to the Lagrangian $\mathcal{L}(F,G)$ by the relation 
\begin{equation}
 \mathcal{H}(P,Q)= 2F\mathcal{L}_{F}(F,G) - \mathcal{L},
\end{equation}
and the Lagrangian is dependent on the invariant formed with 
the Maxwell tensor $F^{\mu \nu}$.  Plebanski nonlinear 
electrodynamics has been used to study nonlinear optics and 
condensed matter physics. Plebanski theory has been studied 
extensively in the context of gravitational theories and 
regular nonrotating black hole solution has been obtained 
\cite{Ayon-Beato:1998hmi,Ayon-Beato:1999qin,
Ayon-Beato:1999kuh,Ayon-Beato:2000mjt,Ayon-Beato:2004ywd}. Very 
recently, a charged rotating black hole solution using Plebanski 
the theory is also obtained in Refs. 
\cite{Garcia-Diaz:2021bao,DiazGarcia:2022jpc,Ayon-Beato:2022dwg}. Using 
Plebanski nonlinear electrodynamics formalism is an interesting 
model of nonlinearly charged AdS black hole was obtained in 4D 
critical gravity \cite{alvarez2022thermodynamics} and its 
logarithmic corrections to thermodynamics are not studied.  
This provides us with an opportunity to fill this gap. 
 
This work considers an AdS black hole in four-dimensional critical gravity coupled with nonlinear electrodynamics and discusses their thermal properties. Furthermore, we study the effects of thermal fluctuations on the thermodynamics of this black hole. In this regard, we compute first-order correction to the entropy of nonlinearly
charged AdS black holes. Next, we plot the entropy as a function of horizon radius for both cases with
and without considering thermal fluctuations. Here, we find that the thermal fluctuations affect the entropy of
small black holes significantly and for large black holes their impacts are negligible. Moreover, we compute the
corrected mass (enthalpy) of the system using the Hawking temperature and corrected entropy. The pressure
can be expressed in terms of the cosmological constant. So, the conjugate (corrected) thermodynamic volume of
a black hole is calculated using the expression of corrected mass. This can be done based on the fact that the
system must satisfy the first-law of thermodynamics. Once we have expressions of the corrected mass, volume, and entropy, it is a matter of calculation to compute corrected Helmholtz and Gibbs free energy. Here, we find
that the Helmholtz free energy decreases with horizon radius and the thermal fluctuations do not change the
nature of Helmholtz free energy. Thermal fluctuation decreases the Helmholtz free energy a bit. In
the case of Gibbs free energy, we find that for large black holes, Gibbs free energy takes a negative value. Also, we
notice that the effects of thermal fluctuation are significant for small black holes.

The stability of this black hole system is also studied. For this, we calculate the specific heat. The positive
value of specific heat suggests that the black hole is the stable state for the system in equilibrium. However,
due to small statistical fluctuation, the system undergoes to an unstable state for the small black holes. The thermal fluctuations do not affect the stability of large black holes. Next, we compute the effects of
thermal fluctuations on specific heat. The effects of thermal fluctuations on isothermal compressibility are also
computed. A phase transition occurs for the corrected isothermal compressibility from a positive to a negative
value at a critical horizon radius. We also compute the speed of sound for this black hole whose value ranges
from zero to one. Finally, we consider the system as a Van der Waals fluid and it is observed that the pressure is
discontinuous with respect to a specific volume of the black hole.

The main aim of this paper is to study correction on various thermodynamics parameters of nonlinearly
charged AdS black holes in 4D critical gravity when small statistical fluctuations around its equilibrium are
taken into account. In section \ref{sec:2}, we study the black hole solution in critical gravity coupled with nonlinear electrodynamics. Within this section, we study uncorrected electric charge, electric potential, Hawking temperature, Wald entropy, the mass of the black hole, the thermodynamic volume of the black hole, free energy, and specific
heat. In section \ref{sec:3}, we compute the effects of thermal fluctuations on various thermodynamic parameters. In
section \ref{sec:4}, we study the charged AdS black hole in nonlinear electrodynamics as a Van der Waals fluid. Finally,
in section \ref{sec:5}, we summarize our results.

\section{The metric and thermodynamics}\label{sec:2}
In this section, we recapitulate some of the known facts about nonlinear electrodynamics in critical gravity.
Let us begin by writing an action describing the theory of critical gravity coupled with nonlinear electrodynamics \cite{alvarez2022thermodynamics}

\begin{equation}\label{eq:2}
    S[g_{\mu \nu} , A_{\mu}, P^{\mu \nu}]= \int d^{4}x \sqrt{-g} [\mathcal{L}_{CG} + \mathcal{L}_{NLE}] ,
\end{equation}
where $\mathcal{L}_{CG}$ and $\mathcal{L}_{NLE}$ are the Lagrangian of the critical gravity and nonlinear electrodynamics, respectively. Here, $\mathcal{L}_{CG}$ has the following form:
\begin{equation}\label{eq:3}
    \mathcal{L}_{CG}= \frac{1}{2 \kappa} \Bigl( R - 2 \Lambda + \beta_{1} R^{2} +\beta_{2} R_{\mu \nu} R^{\mu \nu}  \Bigl) ,
\end{equation}
where $\kappa$ is the surface gravity, $R$ and $R_{\mu \nu}$ are the Ricci scalar and Ricci tensor, $\Lambda$ is the cosmological constant, $\beta_1$
and $\beta_2$ are the coupling constants. Critical gravity allows for the massive spin-zero fields to vanish if the coupling constants $\beta_1$ and $\beta_2$ are restricted to obey the relations \cite{alvarez2022thermodynamics}
\begin{equation}\label{eq:4}
\begin{split}
\beta_{2} &= -2\beta_1, \\
\beta_1 &=-\frac{1}{2\Lambda}.
\end{split}
\end{equation}
The expression for the Lagrangian describing the nonlinear electrodynamics is given by 
\begin{equation}\label{eq:5}
    \mathcal{L}_{NLE} = -\frac{1}{2} P^{\mu \nu} F_{\mu \nu} + \mathcal{H}(P,Q),
\end{equation}
where $P^{\mu \nu}$ is conjugate antisymmetric tensor known as Plebanski tensor and  structure-function $\mathcal{H}(P,Q)$,  where $P$ and
$Q$ are the invariants formed with the antisymmetric conjugate Plebanski tensor. The field strength tensor $F_{\mu \nu}$ is defined in terms of vector field $A_{\mu}$ as: $F_{\mu \nu} = \partial_{\mu}A_{\nu} - \partial_{\nu}A_{\mu}$. $\mathcal{H}(P)$ is  a structure-function depending on the invariant formed with the conjugated antisymmetric tensor.

Variation of the action \eqref{eq:2}  gives the following field equations:
\begin{equation}\label{eq:6}
\begin{split}
G_{\mu \nu} + \Lambda g_{\mu \nu} + \chi_{\mu \nu}^{CG} &=\kappa T_{\mu \nu}^{NLE}, \\
\nabla_{\mu} P^{\mu \nu} &=0,
\end{split}    
\end{equation}
where 
\begin{eqnarray}
     \chi_{\mu \nu}^{CG}&= &2 \beta_{2} \Bigl( R_{\mu \rho} R_{\nu}^{ \rho} - \frac{1}{4} R^{\rho \sigma} R_{\rho \sigma} g_{\mu \nu} \Bigl) + 2 \beta_{1} R \Bigl( R_{\mu \nu} - \frac{1}{4} R g_{\mu \nu}  \Bigl) + \beta_{2} \Bigl( \Box R_{\mu \nu} \nonumber\\
  & +& \frac{1}{2} \Box  g_{\mu \nu} -2 \nabla_{\rho} \nabla_{( \mu} R_{\nu )}^{\rho} \Bigl) +2 \beta_{1} \Bigl( g_{\mu \nu} \Box R - \nabla_{\mu} \nabla_{\nu} R \Bigl) ,\label{eq:7} \\
     T_{\mu \nu}^{NLE} &=& \mathcal{H}_{p} P_{\mu \lambda} P_{\nu}^{\lambda} - g_{\mu \nu} \bigl( 2 P \mathcal{H}_{p} -\mathcal{H}  \bigl),\label{eq:8}
\end{eqnarray}
where $\beta_1$ and $\beta_2$ are given in equation \eqref{eq:4} and    $
 \mathcal{H}_{P}   =\frac{\partial{\mathcal{H}}}{\partial{P}}$.  
The  asymptotically AdS black hole metric is given by \cite{alvarez2022thermodynamics}
\begin{equation}\label{eq:10}
    ds^{2}= - \frac{r^{2}}{l^{2}} f(r) dt^{2} + \frac{l^{2}}{r^{2}} \frac{ dr^{2}}{f(r)} + \frac{r^{2}}{l^{2}} d{\Omega_{2}}^{2},
\end{equation}
where cosmological constant $\Lambda= - \frac{3}{l^2}$, with the  asymptotic condition
\begin{equation}\label{eq:11}
 \lim_{r \to \infty} f(r) =1.
\end{equation}
The nonlinear source is described by the structure-function $\mathcal{H}$ is real. Here we choose the structure-function depends only $P$, because we are interested in static configurations and $\mathcal{H}=\mathcal{H}(P)$ \cite{alvarez2022thermodynamics} 
\begin{equation}\label{eq:12}
    \mathcal{H}(P) = \frac{(\alpha_{2}^{2} - 3 \alpha_{1} \alpha_{3} ) l^{2} P}{3 \kappa} - \frac{2 \alpha_{2} (-2P)^{\frac{1}{4}}}{l \kappa} + \frac{\alpha_{2} \sqrt{-2P}}{\kappa},
\end{equation}
 where $\alpha_{1}$, $\alpha_{2}$, and $\alpha_{3}$ are coupling constant. From second equation of \eqref{eq:6} one can obtain 
 \begin{equation}\label{eq:13}
     P=-\frac{M^2}{2r^4},
 \end{equation}
where $M$ is  an integration constant. Therefore, the 
structure-function $\mathcal{H}$ in equation \eqref{eq:12} is real. Finally, using field equation \eqref{eq:6} one can obtain the function $f(r)$ as 
\begin{equation}\label{eq:14}
    f(r) = 1 - \alpha_{1} \sqrt{M} \frac{l}{r} +  \alpha_{2} {M} \frac{l^2}{r^2} - \alpha_{3} {M}^{\frac{3}{2}} \frac{l^3}{r^3}.
\end{equation}
It is shown in Ref. \cite{alvarez2022thermodynamics} that the structural coupling constants have a significant role in the characterization of
the solutions.

\section{Thermodynamics}\label{sec:3}
In this section, we study the thermodynamics of nonlinearly charged AdS black holes in critical gravity. The metric of such a black hole is given in equation \eqref{eq:14}. The electric charge of the black hole is calculated by \cite{alvarez2022thermodynamics}
\begin{equation}\label{eq:15}
    Q= \frac{\Omega_{2} r_{h}^{2}}{\zeta^{2} l^4},
\end{equation}
where $r_{h}$ is the position of the horizon and  can be expressed as $ r_{h} = \zeta \sqrt{M} l$, where $\zeta$ is the roots of the  polynomial
 $\zeta^{3} - \alpha_{1} \zeta^{2} + \alpha_{2} \zeta - \alpha_{3} =0$. The electric potential is 
\begin{equation}\label{eq:16}
    \Phi= \frac{r_{h}}{\kappa} \bigl( \alpha_{2} + \alpha_{1}^{2}  - \frac{3}{2} \alpha_{1} \zeta - \frac{\alpha_{1} \alpha_{2}}{\zeta} +\frac{\alpha_{2}^{2}}{3 \zeta^{2}}   \bigl).
\end{equation}
 The Hawking temperature  due to surface gravity is calculated by
\begin{equation}\label{eq:17}
    T_{H} = \frac{r_{h}}{4 \pi l^2}  \bigl( 3 - \frac{2 \alpha_{1}}{\zeta} + \frac{\alpha_{2}}{\zeta^{2}}  \bigl).
\end{equation}
The Wald entropy is given by \cite{wald1993black,alvarez2022thermodynamics}
\begin{equation}\label{eq:18}
    S_{0}= \frac{2 \Omega_{2} \pi}{\kappa} \Bigl( \frac{r_{h}}{l} \Bigl)^{2} \biggl( \frac{\alpha_{1}}{\zeta} - \frac{2 \alpha_{2}}{3 \zeta^{2}} \biggl),
\end{equation}
where $\Omega_2$ refers to the finite volume of the compact planar manifold.
The mass of the black hole in central gravity with nonlinear electrodynamics has the following expression \cite{alvarez2022thermodynamics}:
\begin{equation}\label{eq:19}
    \mathcal{M} = \frac{\alpha_{1} \alpha_{2} \Omega_{2} r_{h}^{3}}{9 \kappa \zeta^{3} l^4}=\frac{64 \alpha_{1} \alpha_{2} \Omega_{2} \pi^{2} P^{2} r_{h}^{3}}{81 \kappa \zeta^{3}},
\end{equation}
This is equivalent to the enthalpy of the system. It is well known that the cosmological constant is responsible for pressure in AdS space, $P=-\Lambda/8 \pi= 3/8 \pi l^2$. Now, the mass of the black hole in terms of Wald entropy and charge written by
\begin{equation}\label{eq:20}
    \mathcal{M}(S_0,Q) =\frac{\sqrt{6 \kappa} S_{0}^{3/2} \bigl(  3\zeta^{2} -2 \alpha_{1} \zeta + \alpha_{2} \bigl)}{12 \sqrt{\Omega_{2}} \pi^{3/2} \sqrt{3 \alpha_{1} \zeta -2 \alpha_{2}} l \zeta}  + \frac{Q^{3/2} l^2 \zeta \Psi }{9 \sqrt{\Omega_{2}} \kappa},
\end{equation}
where $\Psi$ is given by 
\begin{equation}\label{eq:21}
    \Psi= 6 \alpha_{2} +6 \alpha_1^2 -9\alpha_1 \zeta - \frac{6 \alpha_1 \alpha_2}{\zeta} + \frac{2 \alpha_2^2}{\zeta}.
\end{equation}
The conjugate volume of the black hole is \cite{kubizvnak2017black} 
\begin{equation}\label{eq:22}
    V=\frac{\partial{M}}{\partial{P}}= \frac{128  \alpha_{1} \alpha_{2} \Omega_{2} \pi^{2} P r_{h}^{3}}{81 \kappa \zeta^{3}}.
\end{equation}
With the above-mentioned thermodynamical quantities, we can compute further properties of the black hole such as internal energy $(U)$, Helmholtz free energy $(F)$, and Gibbs free energy $(G)$. The internal energy is
calculated as
\begin{equation}\label{eq:23}
    U= \mathcal{M} - PV= -\frac{\alpha_{1} \alpha_{1} \Omega_{2} r_{h}^{3} }{9 \kappa \zeta^{3} l^4}.
\end{equation}
Using the standard definition of Helmholtz free energy, we obtain
\begin{equation}\label{eq:24}
    F=U-T_{H}S_0 = -\frac{\Omega_{2} r_{h}^{3} (27 \zeta^{3} \alpha_{1} -18 \zeta^{2} \alpha_{1}^{2}-18 \zeta^{2} \alpha_{2} +23 \alpha_{1} \alpha_{2} \zeta -6 \alpha_{2}^{2})}{18 \zeta^{4} \kappa  l^{4}} .
\end{equation}
Now, it is a matter of calculation to obtain the expression of the Gibbs free energy $ G=\mathcal{M} - T_HS_0-\Phi Q$. This reads 
\begin{equation}\label{eq:26}    
    G= -\frac{3 \Omega_{2} r_{h}^{2}}{2 \kappa \zeta^{4} l^4}  \biggl( \alpha_{1} (r_{h} - 1) \zeta^{3} - \frac{2(\alpha_{1}^{2} +\alpha_{2} ) (r_{h} - 1) \zeta^{2}}{3}   +\frac{19 (r_{h} - \frac{18}{19}) \alpha_{1} \alpha_{2} \zeta}{27} - \frac{2 \alpha_{2}^{2} (r_{h} + 1)}{9}
    \biggl).
    \end{equation}
The specific heat of a black hole plays an important role in the stability of the system. Now, we calculate the specific heat at constant electric potential as
\begin{equation}\label{eq:27}
    C_{\Phi} = T_{H}  \biggl(\frac{\partial{S_{0}}}{\partial{T}} \biggl)_{\Phi} ,
\end{equation}
\begin{equation}\label{eq:28}
    C_{\Phi} = \frac{4 \Omega_{2} \pi  r_{h}^{2} (3 \alpha_{1} \zeta -2 \alpha_{2} )}{3 \kappa \zeta^{2}   l^{2}}.
\end{equation}
The temperature and specific heat will be positive if and only if  
\begin{equation*}
    \Psi_1= (3\zeta^{2} -2 \alpha_{1} \zeta + \alpha_{2}) \geq 0
\end{equation*}
\begin{equation}\label{eq:29}
    \Psi_2=(3 \alpha_{1} \zeta -2 \alpha_{2} ) \geq 0.
\end{equation}
If $T_{H},\Phi \geq 0$, then above equation and 
$     \Psi= \frac{\alpha_1 \alpha_2}{\zeta} - \frac{\Psi_1 \Psi_2}{\zeta^2} \geq 0,$
hold, where $\Psi$ is defined in equation \eqref{eq:21}.
\begin{figure}[hbt]
    \centering
    \includegraphics[width=10cm , height=6cm]{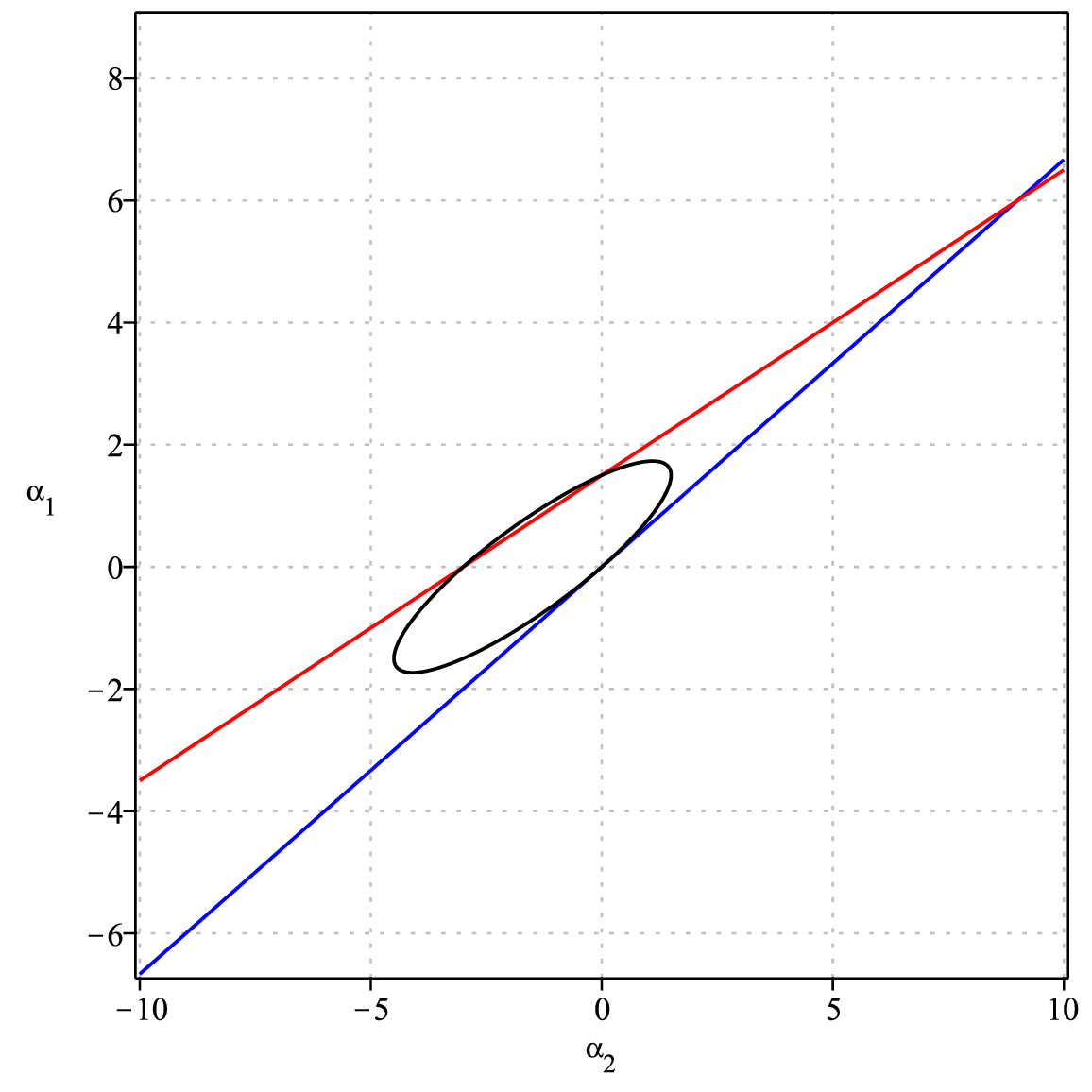}
    \caption{ $\Psi_1$ denoted by red line, $\Psi_2$ denoted by blue line and $\Psi$ denoted by black line with $\zeta=1$.}
    \label{fig:1}
\end{figure}
The possible solutions for $\alpha_1$ and $\alpha_2$ are shown in Fig. \ref{fig:1}. The region bounded by the red and blue curve in the first octant is the possible solution, except the region bounded by the black curve.
\section{ Thermodynamics with first-order correction}\label{sec:4}
In this section, we study the effect of thermal fluctuations on various thermodynamic parameters of black holes up to the first order. To the first order, correction to entropy was first studied in Ref. \cite{das2002general}, which is logarithmic in nature. The logarithmic corrections to thermodynamic entropy arise when small stable fluctuations around equilibrium are taken into account. The logarithmic correction to the entropy of BTZ black hole, Schwarzschild AdS black hole \& Reissner-Nordstrom black hole studied in Ref. \cite{das2002general,upadhyay2022modified}. The partition function of a thermodynamics system is
\begin{equation}\label{eq:31}
    Z(\beta) = \int_{0}^{\infty} \rho(E) e^{-\beta E} dE,
\end{equation}
where $\beta=1/T_{H}$. The density of states for fixed energy can be obtained from the above equation using inverse Laplace transformation
\begin{equation}\label{eq:32}
    \rho= \frac{1}{2 \pi i} \int_{c - \textbf{i} \infty}^{c + \textbf{i} \infty} e^{\mathcal{S}(\beta)} d{\beta}.
\end{equation}
The above complex integral can be computed using the steepest descent method at saddle point $\beta_{0}$, and we obtain
\begin{equation}\label{eq:33}
    \mathcal{S}(\beta)= S_{0} + \frac{1}{2} (\beta - \beta_0)^2 \biggl( \frac{\partial^2 \mathcal{S}_{0}}{\partial \beta^2} \biggl)_{\beta_{0}} + \cdots,
\end{equation}
where $\mathcal{S}(\beta)$ and $S_0$  is the exact entropy and zeroth order entropy. Substituting equation \eqref{eq:33} into equation \eqref{eq:32} we have 
\begin{equation}\label{eq:34}
    \rho= \frac{e^{S_0}}{2 \pi i} \int_{c - \textbf{i} \infty}^{c + \textbf{i} \infty} e^{\frac{1}{2} (\beta - \beta_0)^2 ( \frac{\partial^2 \mathcal{S}_{0}}{\partial \beta^2} )_{\beta_{0}}} d{\beta}.
\end{equation}
Finally, the above integral gives \cite{das2002general}
\begin{equation}\label{eq:35}
    \rho(E)= \frac{e^{S_0}}{\sqrt{2 \pi \mathcal{S}^{\prime \prime}(\beta_{0})}}.
\end{equation}
 Therefore, the exact entropy due to thermal fluctuation is
\begin{equation}\label{eq:36}
    S= \ln{\rho}= S_0 - \frac{1}{2} \ln{\left( \frac{\partial^2 \mathcal{S}_{0}}{\partial \beta^2} \right)}.
\end{equation}
From \cite{das2002general} one can write the above entropy as
\begin{equation}\label{eq:37}
    S = S_{0} - \frac{1}{2} \ln{(S_{0} T_{H}^{2})}.
\end{equation}
Therefore, entropy received correction due to thermal fluctuations. Now, to identify the effect of this correction term on other thermodynamical quantities, we  label the $1/2$ factor in the R.H.S of equation \eqref{eq:37} by $\gamma$. Fianlly, the corrected entropy becomes 
\begin{equation}\label{eq:38}
   S = S_{0} - \gamma \ln{(S_{0} T_{H}^{2})},   
\end{equation}
where $\gamma=0$ refers to uncorrected entropy $S_0$ and $\gamma=1/2$ refers to corrected entropy in equation \eqref{eq:37}. Therefore, the correction coefficient $\gamma$ can only take two values, i.e. $\gamma=0$ or $0.5$ \& $\gamma$ is a dimensionless quantity. The correction coefficients $\gamma$ arise due to the thermal fluctuation in the equilibrium thermodynamics of black holes. This thermal nature leads to a prefactor in the expression for the density of states of the system, which in turn modifies the entropy of the black hole. 
\subsection{Corrected entropy}
\begin{figure}[hbt]
    \centering
    \includegraphics[width=10cm , height=6cm]{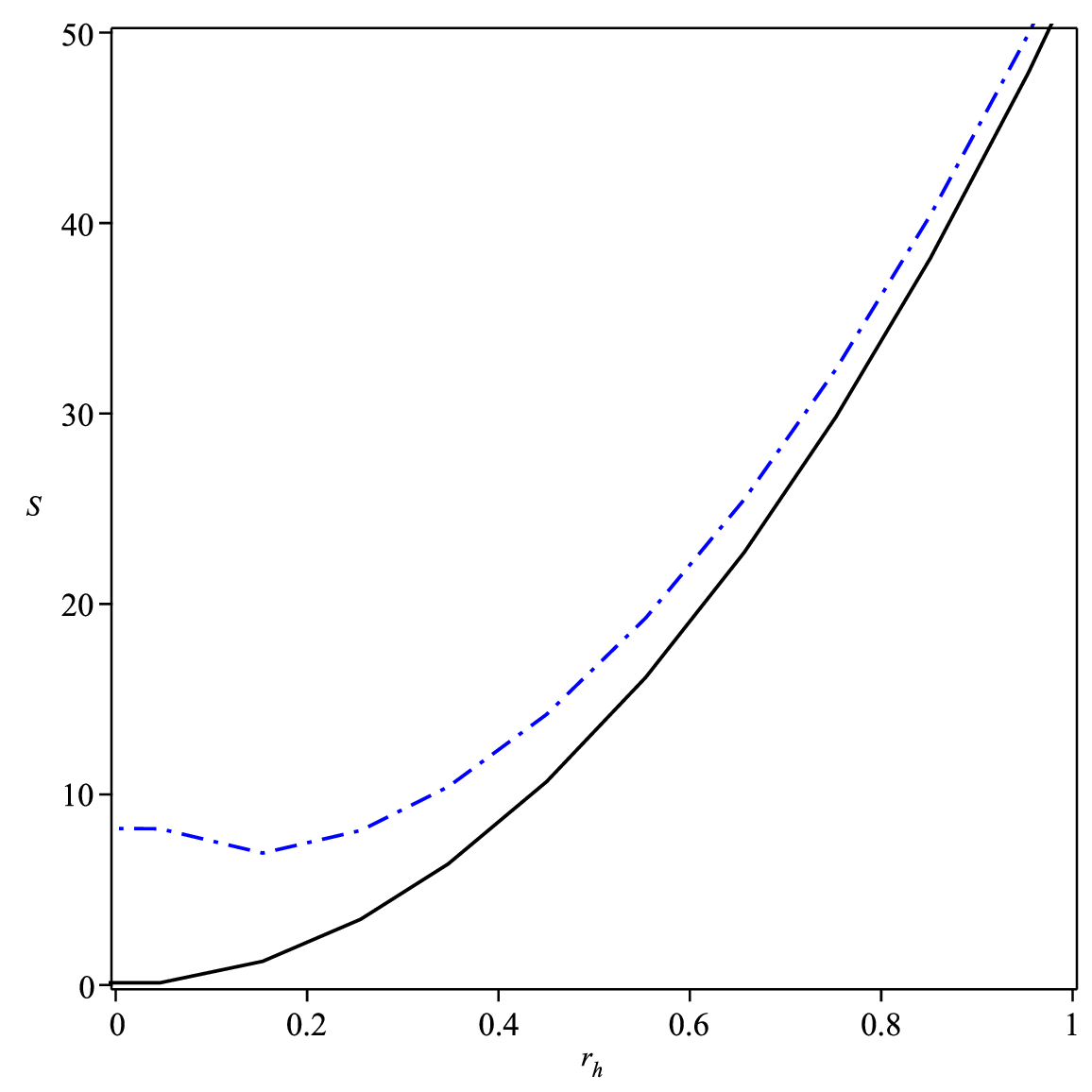}
    \caption{ Entropy vs. black hole horizon with $\zeta=1$, $\alpha_{1} = \alpha_{2}=2$ and $l=2$. Here $\gamma=0$ is denoted by a black line and $\gamma=0.5$ is denoted by a blue dash-dot line.}
    \label{fig:2}
\end{figure}

Using the relations    \eqref{eq:17}, \eqref{eq:18} and  \eqref{eq:38},  we obtain 
\begin{equation}\label{eq:39}
    S = \frac{2 \Omega_{2} \pi  r_{h}^{2} (\alpha_{1} \zeta -\frac{2 \alpha_{2}}{3})}{\zeta^{2} \kappa  l^{2}}-\gamma  \ln  \Biggr[ \frac{\Omega_{2} r_{h}^{4} (\alpha_{1} \zeta -\frac{2 \alpha_{2}}{3}) (3 \zeta^{2}-2 \alpha_{1} \zeta +\alpha_{2} )^{2}}{8 \pi  \zeta^{6} \kappa  l^{6}}\Biggr].
\end{equation}
The effects of thermal fluctuation on entropy are depicted in Fig. \ref{fig:2}. From the diagram \ref{fig:2}, we observe that entropy increases with the horizon radius. The thermal fluctuation increases the entropy of the black hole significantly for small-sized black holes with horizon radius $r_h < 0.2$. Therefore, quantum effects significantly dominate for smaller-sized black holes with $r_h < 0.2$. However, for larger black holes with $r_h>0.2$, the effects of thermal fluctuations on the entropy are negligible.

\subsection{Corrected mass}
Now, we analyze the effect of thermal fluctuation on the total mass (enthalpy) of the black holes. The corrected mass can be evaluated with the help of the following definition:
\begin{equation}\label{eq:40}
    \mathcal{M}_{c} = \int T_{H} dS.
\end{equation}
Here, we have introduced the corrected entropy in place of equilibrium entropy. Substituting the value of Hawking temperature and corrected entropy from equations \eqref{eq:17} and \eqref{eq:39}, respectively, into equation \eqref{eq:40}, we have
\begin{equation}\label{eq:41}
    \mathcal{M}_{c} = \frac{(3 \zeta^{2}-2 \alpha_{1} \zeta +\alpha_{2} ) r_{h} (3 \pi  \zeta  r_{h}^{2} \Omega_{2} \alpha_{1} -9 \gamma  \zeta^{2} \kappa  l^{2}-2 \pi  r_{h}^{2} \Omega_{2} \alpha_{2} )}{9 l^{4} \zeta^{4} \pi  \kappa}.
\end{equation}
\begin{figure}[hbt]
    \centering
    \includegraphics[width=10cm , height=6cm]{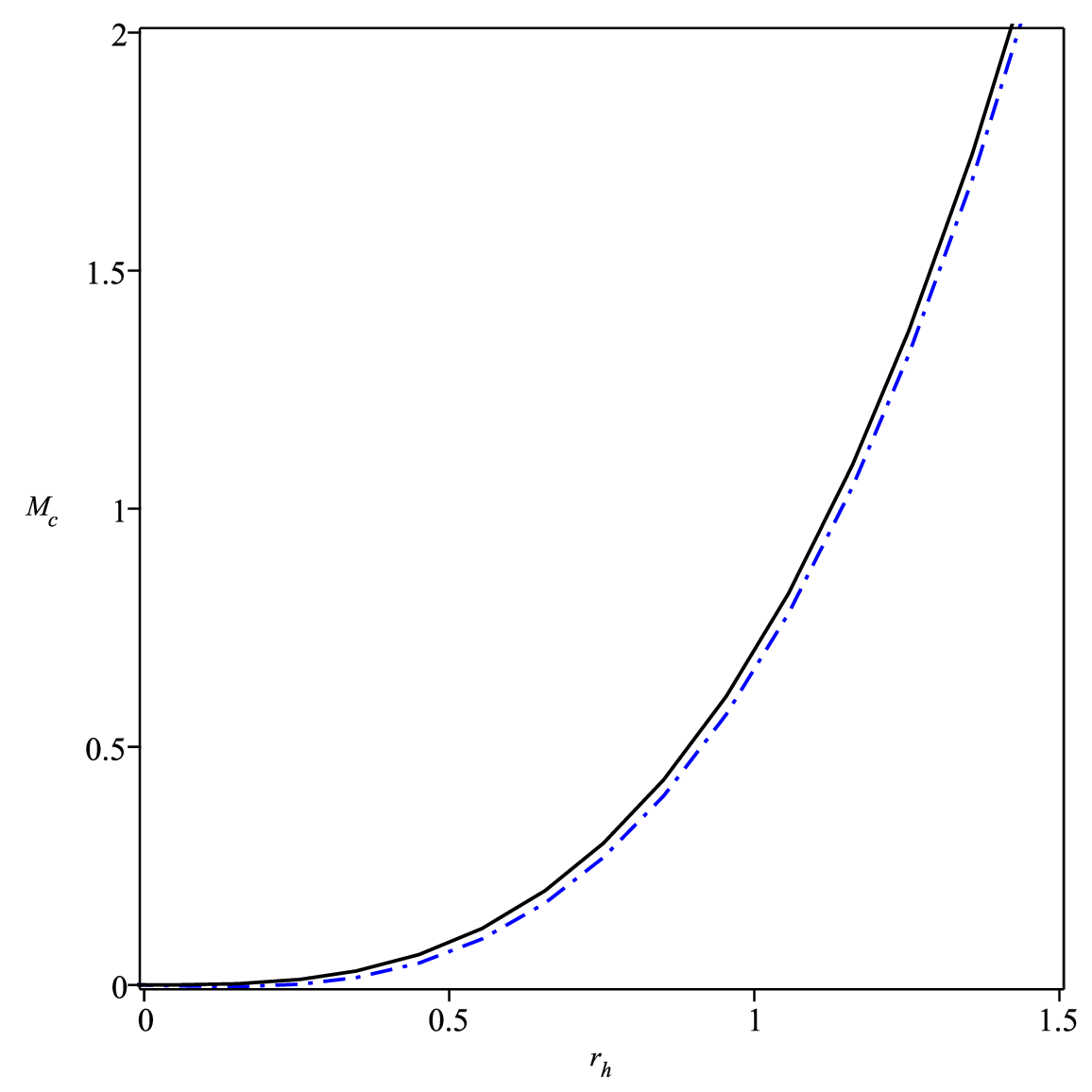}
    \caption{ Mass(enthalpy) Vs.  black hole horizon with $\zeta=1$, $\alpha_{1} = \alpha_{2}=2$ and $l=2$. Here $\gamma=0$ is denoted by a black line and $\gamma=0.5$ is denoted by a blue dash-dot   line.}
    \label{fig:3}
\end{figure}
Now, to do a comparative analysis, we plot the corrected mass and equilibrium mass with respect to the horizon radius in Fig. \ref{fig:3}. From Fig. \ref{fig:3}, we see that the mass is an increasing function of the horizon radius. Interestingly, we find that the thermal fluctuations decrease the mass a bit but do not change the behavior of the mass.
\subsection{Corrected thermodynamic volume of black hole}
\begin{figure}[hbt]
    \centering
    \includegraphics[width=10cm , height=6cm]{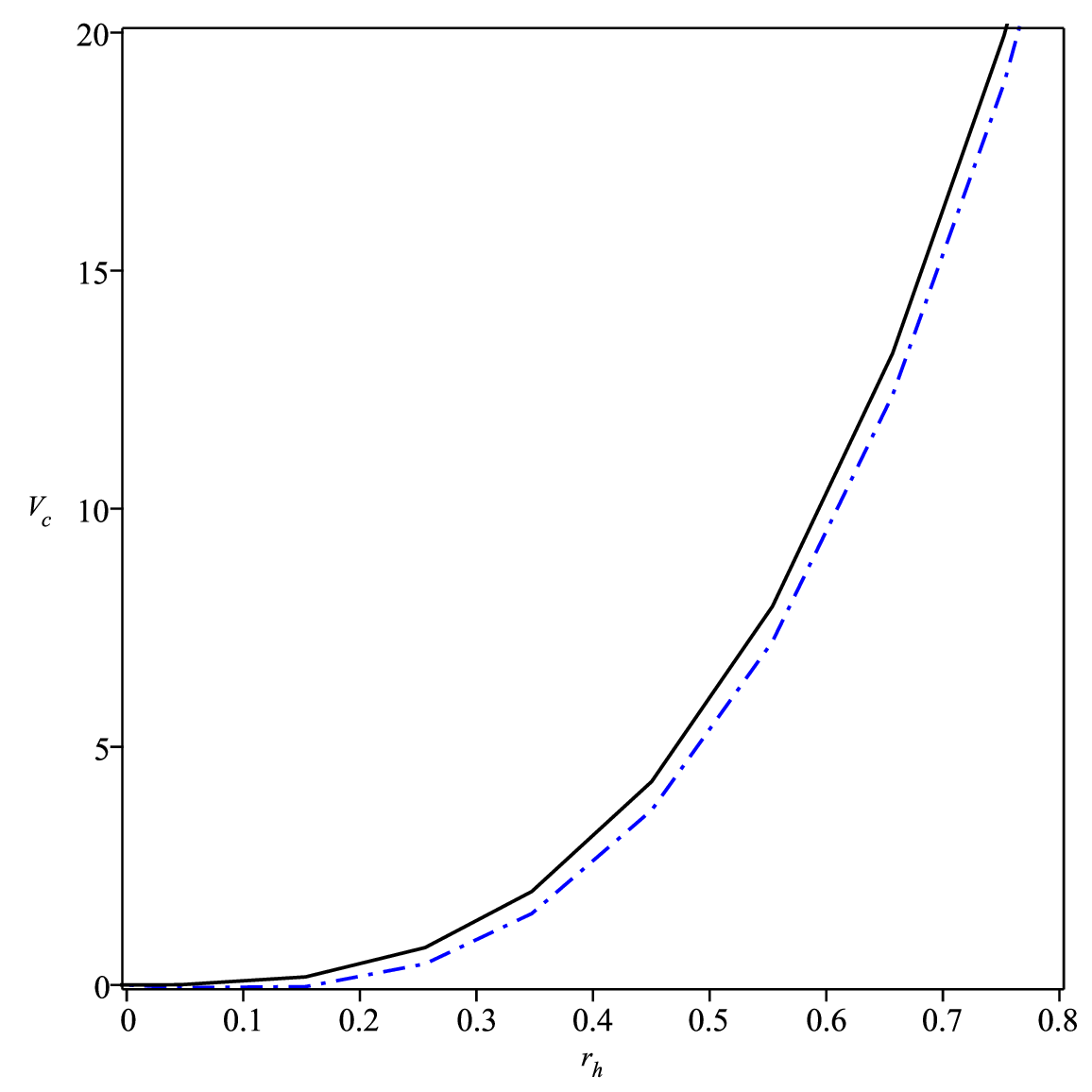}
    \caption{Thermodynamic volume of the black hole Vs.  black hole horizon with $\zeta=1$, $\alpha_{1} = \alpha_{2}=2$ and $l=2$. Here $\gamma=0$ is denoted by the black line, and $\gamma=0.5$ is denoted by the blue dash-dot line.}
    \label{fig:4}
\end{figure}

In this subsection, we study the corrected thermodynamic volume of the black hole as a function of pressure and horizon radius. The thermodynamic volume of an asymptotically AdS black hole is defined as \cite{kubizvnak2017black}
\begin{equation}\label{eq:42}
    V=\biggl( \frac{\partial{\mathcal{M}_{c}}}{\partial{P}} \biggl)_{S_{0},Q},
\end{equation}
To compute this equation, we first write the corrected mass in terms of pressure. Since pressure depends on the cosmological constant. So, the corrected mass  \eqref{eq:41} can be expressed in terms of pressure as follows:
 \begin{equation}\label{eq:43}
    \mathcal{M}_{c} = \frac{64 (3 \zeta^{2}-2 \alpha_{1} \zeta +\alpha_{2} ) r_{h} (3 \pi  \zeta  r_{h}^{2} \Omega_{2} \alpha_{1} -\frac{27 \gamma  \zeta^{2} \kappa}{8 \pi  P}-2 \pi  r_{h}^{2} \Omega_{2} \alpha_{2} ) \pi  P^{2}}{81 \zeta^{4} \kappa}.
\end{equation}
Substituting the value of \eqref{eq:43} in Eq. \eqref{eq:42}, we obtain the corrected thermodynamic volume of black hole as
\begin{equation}\label{eq:44}
    V_{c}=\frac{8 (3 \zeta^{2}-2 \alpha_{1} \zeta +\alpha_{2} ) r_{h} (48 P \pi^{2} \zeta  r_{h}^{2} \Omega_{2} \alpha_{1} -32 P \pi^{2} r_{h}^{2} \Omega_{2} \alpha_{2} -27 \gamma  \zeta^{2} \kappa )}{81 \zeta^{4} \kappa},
\end{equation}
This further simplifies to   
\begin{equation}\label{eq:45}
    V_{c}=\frac{8 (3 \zeta^{2}-2 \alpha_{1} \zeta +\alpha_{2} ) r_{h} (6 \pi  \zeta  r_{h}^{2} \Omega_{2} \alpha_{1} -9 \gamma  \zeta^{2} \kappa  l^{2}-4 \pi  r_{h}^{2} \Omega_{2} \alpha_{2} )}{27 l^{2} \zeta^{4} \kappa}.
\end{equation}
To study the behavior of thermodynamic volume and their dependency on thermal fluctuations, we plot Fig. \ref{fig:4}. We find that the thermodynamic volume of the black hole increases with the horizon radius. The thermal fluctuation decreases the volume of the black holes which becomes significant in larger black holes.

\subsection{Corrected Helmholtz free energy}
In this subsection, we study the corrected Helmholtz free energy of the AdS black hole due to thermal fluctuation. The Helmholtz free energy is defined by
\begin{equation}\label{eq:46}
     F_{c}= U - T_{H} S.
\end{equation}
By plugging the values from Eqs. \eqref{eq:17}, \eqref{eq:23} and \eqref{eq:39}, the above expression leads to
\begin{equation}\label{eq:47}
    F_{c}= \frac{r_{h} (3 \zeta^{2}-2 \alpha_{1} \zeta +\alpha_{2} )  \biggr[9 \gamma  \ln  \biggl(\frac{\Omega_{2} r_{h}^{4} (3 \alpha_{1} \zeta -2 \alpha_{2} ) (3 \zeta^{2}-2 \alpha_{1} \zeta +\alpha_{2} )^{2}}{24 \pi  \zeta^{6} \kappa  l^{6}}\biggl) \zeta^{2} \kappa  l^{2}-30 \pi  \zeta  r_{h}^{2} \Omega_{2} \alpha_{1} +20 \pi  r_{h}^{2} \Omega_{2} \alpha_{2}  \biggr]}{36 l^{4} \zeta^{4} \pi  \kappa}.
\end{equation}
\begin{figure}[hbt]
    \centering
    \includegraphics[width=10cm , height=6cm]{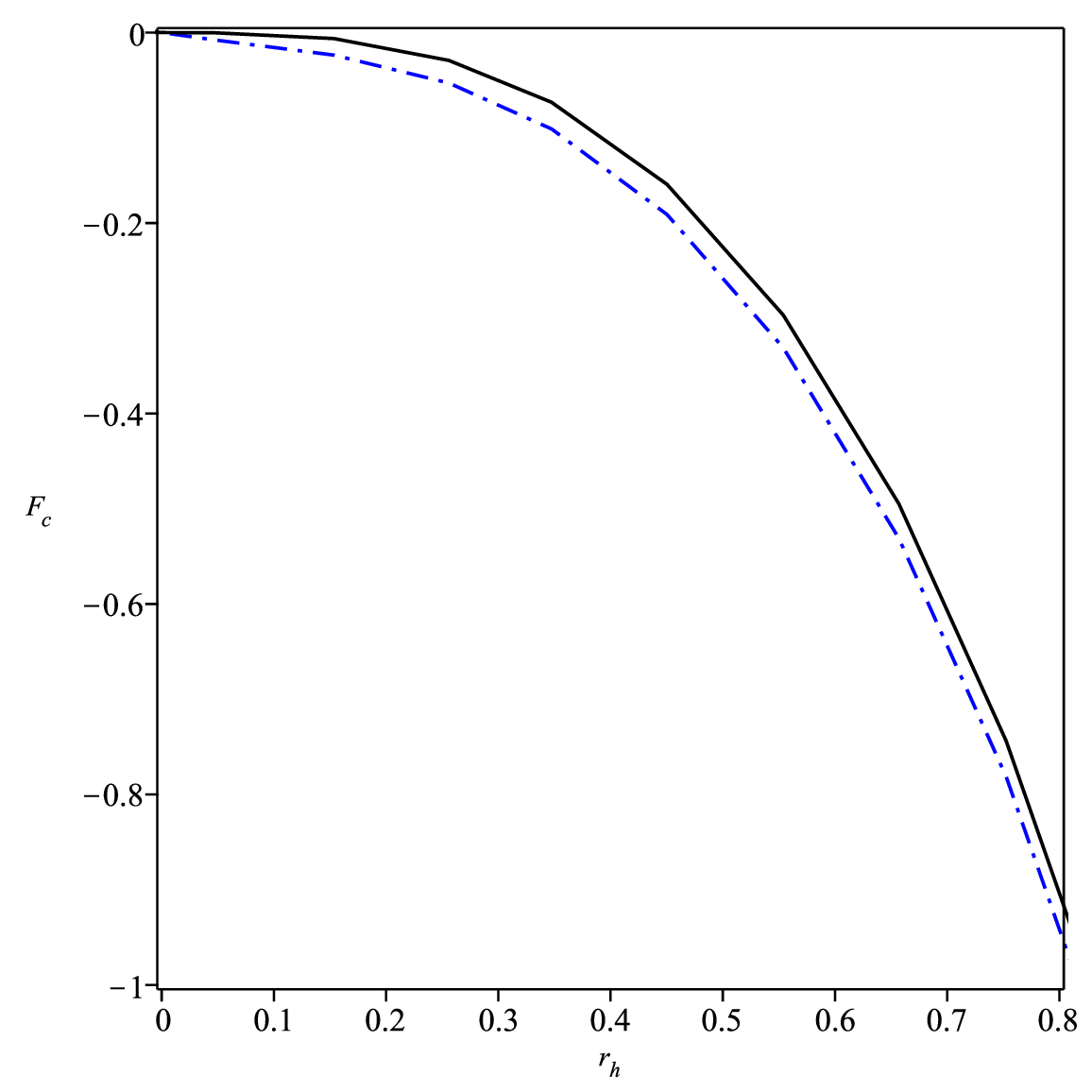}
    \caption{Helmholtz free energy  Vs.  black hole horizon with $\zeta=1$, $\alpha_{1} = \alpha_{2}=2$ and $l=2$. Here $\gamma=0$ is denoted by a black line and $\gamma=0.5$ is denoted by a blue dash-dot line.}
    \label{fig:5}
\end{figure}
To study the behavior of Helmholtz free energy and their dependencies on thermal fluctuation, we plot Fig. \ref{fig:5}. From the figure, it is evident that the Helmholtz free energy decreases with the horizon radius. The thermal fluctuation does not change the nature of Helmholtz free energy. Thermal fluctuation decreases the Helmholtz free energy a bit.
 
\subsection{Corrected Gibbs free energy}
Another important thermal quantity that plays important role in the discussion of the stability of black holes is Gibbs free energy. The Gibbs free energy is defined by
\begin{equation}\label{eq:48}
    G_{c} = \mathcal{M}_{c} -T_{H}S_{c} - {\Phi} Q.
\end{equation}
Substituting the expression of $\mathcal{M}_{c}$, $T_{H}$, $S$, $\Phi$ and $Q$ from equations \eqref{eq:41}, \eqref{eq:17}, \eqref{eq:39}, \eqref{eq:16} and  \eqref{eq:15} into above equation, we have
\begin{eqnarray}
    G_{c} &=& -\frac{\biggl(-9 r_{h} \alpha_{1} \zeta^{3}+6 r_{h} (\alpha_{1}^{2}+\alpha_{2} ) \zeta^{2}-6 r_{h} \alpha_{1} \alpha_{2} \zeta -2 r_{h} \alpha_{2}^{2}\biggl) \Omega_{2} r_{h}}{6 \kappa  \zeta^{4} l^{4}}\nonumber\\
    &-&\frac{ \biggl(3 \zeta^{2}-2 \alpha_{1} \zeta +\alpha_{2} \biggl) \Biggr[\frac{2 \Omega_{2} \pi  r_{h}^{2} (\alpha_{1} \zeta -\frac{2 \alpha_{2}}{3})}{\zeta^{2} \kappa  l^{2}}-\gamma  \ln  \biggl(\frac{\Omega_{2} r_{h}^{4} (\alpha_{1} \zeta -\frac{2 \alpha_{2}}{3}) (3 \zeta^{2}-2 \alpha_{1} \zeta +\alpha_{2} )^{2}}{8 \pi  \zeta^{6} \kappa  l^{6}}\biggl)\Biggr] r_{h}}{4 l^{2} \zeta^{2} \pi}\nonumber\\
   & +&\frac{\biggl(3 \zeta^{2}-2 \alpha_{1} \zeta +\alpha_{2} \biggl)  \biggl(3 \pi  \zeta  r_{h}^{2} \Omega_{2} \alpha_{1} -9 \zeta^{2} \gamma  \kappa  l^{2}-2 \pi  r_{h}^{2} \Omega_{2} \alpha_{2} \biggl) r_{h} }{9 l^{4} \zeta^{4} \pi  \kappa}.\label{eq:49}
\end{eqnarray}
 \begin{figure}[hbt]
\centering
\subfloat[]{\includegraphics[width=.5\textwidth]{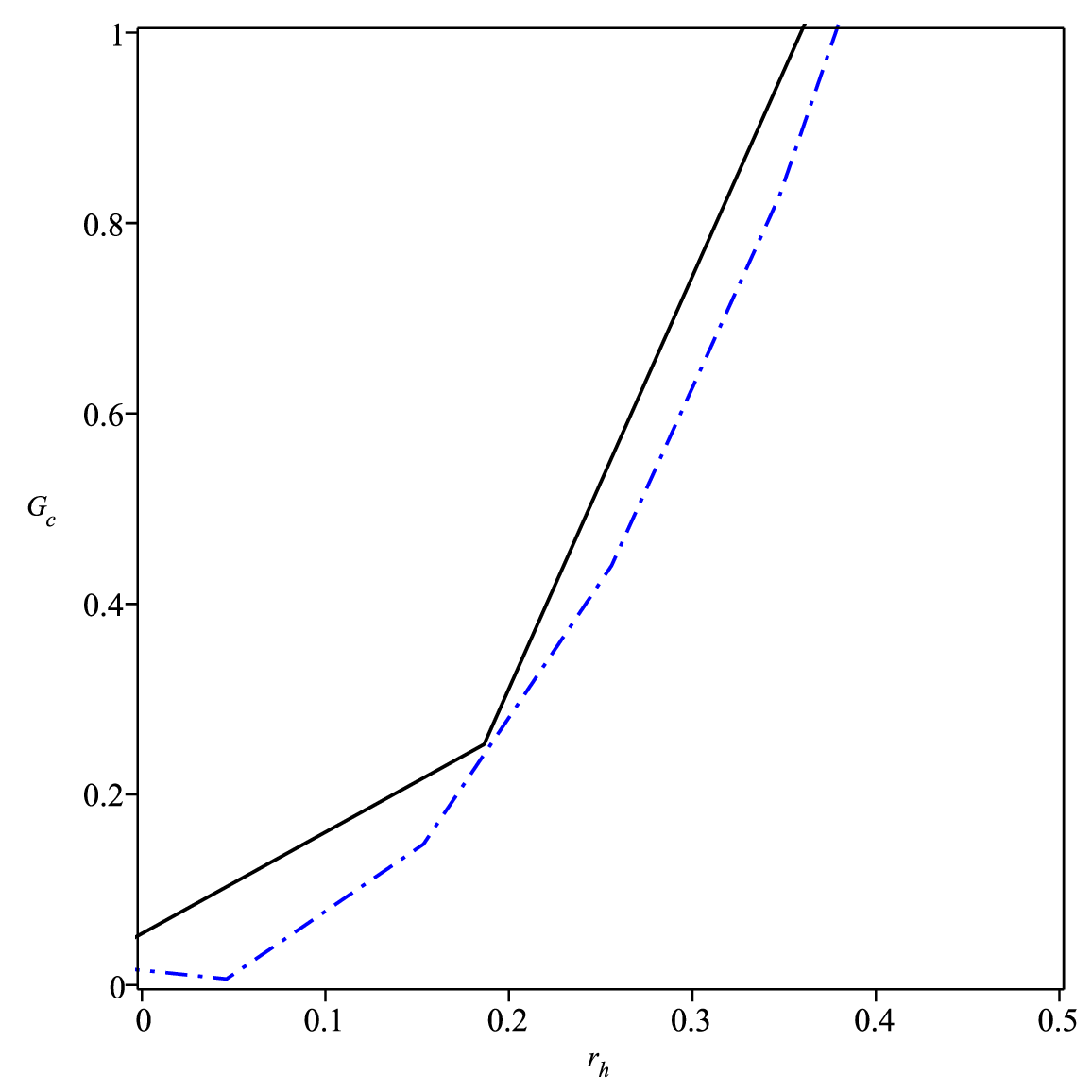}}\hfill
\subfloat[]{\includegraphics[width=.5\textwidth]{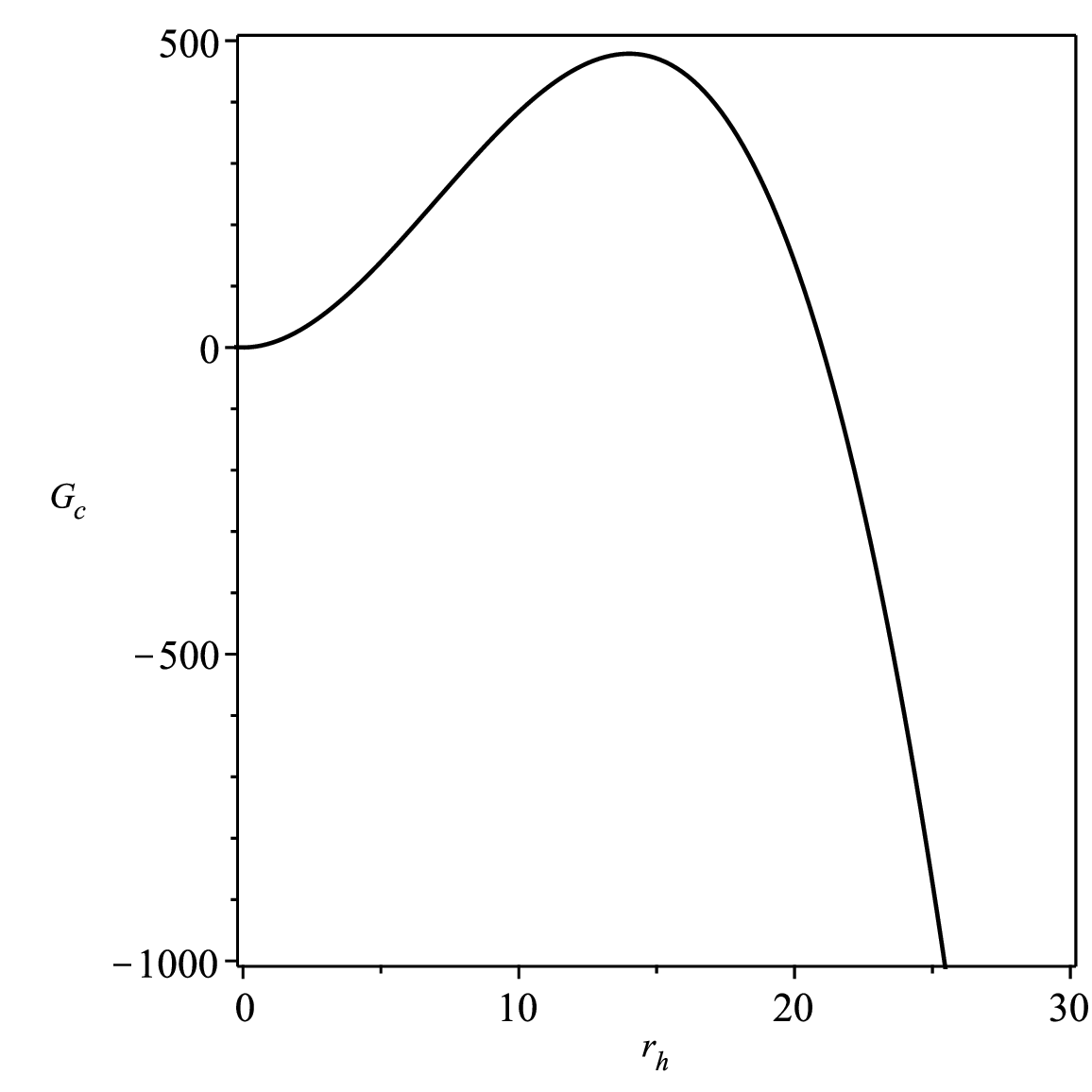}}\hfill
\caption{Gibbs free energy  Vs.  black hole horizon with $\zeta=1$, $\alpha_{1} = \alpha_{2}=2$ and $l=2$. Here $\gamma=0$ is denoted by a black line and $\gamma=0.5$ is denoted by a blue dash-dot line.}\label{fig:6}
\end{figure}
To do a comparative analysis of corrected Gibbs free energy with its equilibrium value, we plot a diagram \ref{fig:6}. The effect of the correction terms 
is significant for small black holes which are depicted in Fig. \ref{fig:6} (a). However, the behavior of Gibbs free energy for the large horizon radius is denoted in
Fig. \ref{fig:6} (b).
From the diagram, it is obvious that Gibbs free energy starts from zero and takes the maximum positive corrected Gibbs free energy with $\gamma=0.5$ value and then starts decreasing towards a negative value. It means that for larger black holes uncorrected Gibbs free energy takes a negative value.

\subsection{Stability and specific heat}

\begin{figure}[hbt]
    \centering
    \includegraphics[width=10cm , height=6cm]{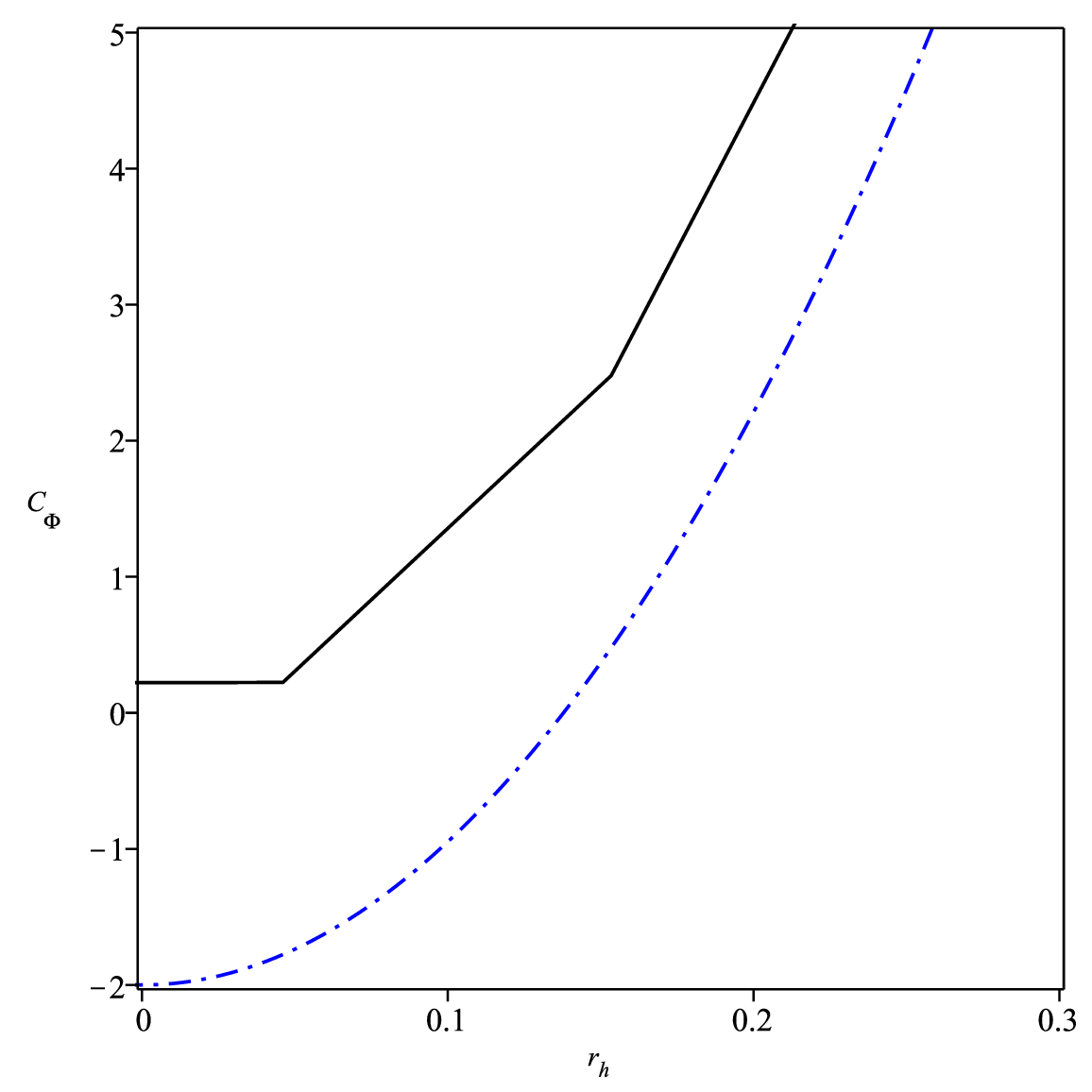}
    \caption{Specific heat Vs.  black hole horizon with $\zeta=1$, $\alpha_{1} = \alpha_{2}=2$ and $l=2$. Here $\gamma=0$ is denoted by a black line and $\gamma=0.5$ is denoted by a blue dash-dot line.}
    \label{fig:7}
\end{figure}

In this subsection, we check the stability of the black hole by estimating the specific heat of the black hole. The specific heat of the black hole is defined by

\begin{equation}\label{eq:50}
   \Bigl( C_{\Phi} \Bigl)_{c} = T_{H} \frac{\partial{S}}{\partial{T_{H}}}=T_{H} \frac{\partial{S}/\partial r_{h}}{\partial{T_{H}}/ \partial r_{h}}.
\end{equation}
Using equations \eqref{eq:17} and \eqref{eq:39}  Specific heat takes the following form:
\begin{equation}\label{eq:51}
    \Bigl( C_{\Phi} \Bigl)_{c} = \biggr[\frac{4 \Omega_{2} \pi  r_{h} (\alpha_{1} \zeta -\frac{2 \alpha_{2}}{3})}{\zeta^{2} \kappa  l^{2}}-\frac{4 \gamma}{r_{h}}\biggr]r_{h}.
\end{equation}
The stability of a black hole is determined by the condition $C_{\Phi} \geq 0 $. Now, we plot the specific heat with respect to the horizon radius to see the signature. From Fig. \ref{fig:7}, we observe that the black hole is stable for $\gamma=0$. However, the thermal fluctuation causes instability to the small ($r_h<0.13$) black holes, i.e. black holes with small horizon radii are thermodynamically locally unstable. A transition from negative specific heat to positive one occurs at $r_h=0.13$ Therefore, black holes with horizon radius $r_h>0.13$ are thermodynamically stable. As the horizon increases thermal fluctuation becomes ineffective to the specific heat, both corrected \& uncorrected specific heat coincide for $r_h >>0.13$.  
\subsection{Corrected isothermal compressibility}
In this subsection, we study the effects of thermal fluctuation on isothermal compressibility and adiabatic compressibility. Let us first define the isothermal compressibility of black hole \cite{dolan2011compressibility}
\begin{equation}\label{eq:52}
    \beta_{T} = -\frac{1}{V_{c}} \biggl( \frac{\partial{V{c}}}{\partial{P}} \biggl)_{T}.
\end{equation}
Using equations \eqref{eq:44} and \eqref{eq:52} isothermal compressibility takes the following form: 
\begin{equation}\label{eq:53}
    \beta_{T} = -\frac{48 \pi^{2} \zeta  r_{h}^{2} \Omega_{2} \alpha_{1} -32 \pi^{2} r_{h}^{2} \Omega_{2} \alpha_{2}}{48 P \pi^{2} \zeta  r_{h}^{2} \Omega_{2} \alpha_{1} -32 P \pi^{2} r_{h}^{2} \Omega_{2} \alpha_{2} -27 \gamma  \zeta^{2} \kappa}. 
\end{equation} 
\begin{figure}[hbt]
    \centering
    \includegraphics[width=10cm , height=6cm]{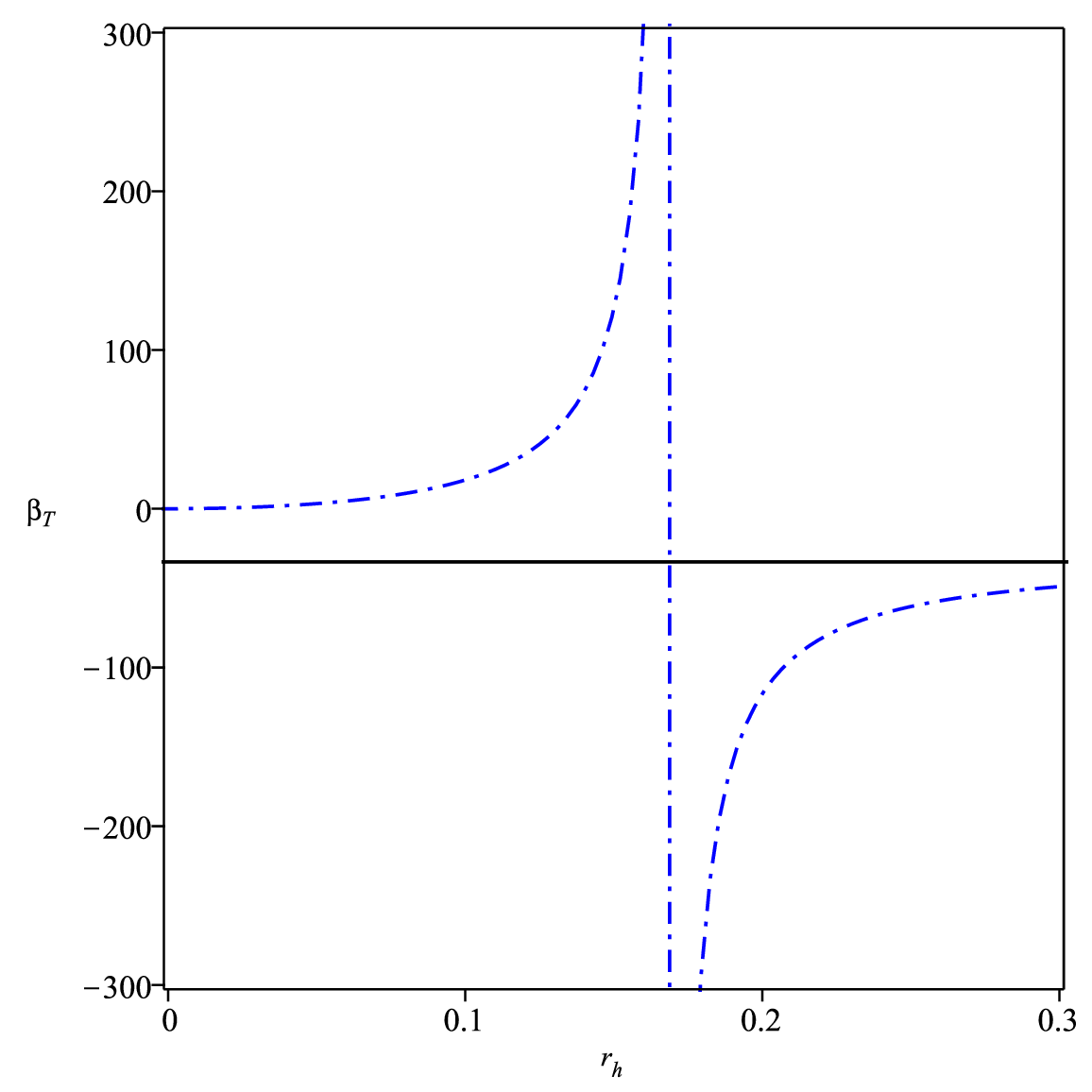}
    \caption{ Isothermal compressibility Vs.  black hole horizon with $\zeta=1$, $\alpha_{1} = \alpha_{2}=2$ and $l=2$. Here $\gamma=0$ is denoted by the black line and $\gamma=0.5$ is denoted by a blue dash-dot line.}
    \label{fig:8}
\end{figure}
Now, we plot the isothermal compressibility of nonlinearly charged AdS black hole in fourth dimensions critical gravity in Fig. \ref{fig:8}. Here, we find that for $\gamma=0$ (equilibrium) isothermal compressibility takes constant
negative value. However, for $\gamma=0.5=0$ (considering thermal fluctuations), a phase transition for the isothermal compressibility occurs which takes a positive value for small-sized black holes and a negative value for massive black holes.

\begin{figure}[hbt]
    \centering
    \includegraphics[width=10cm , height=6cm]{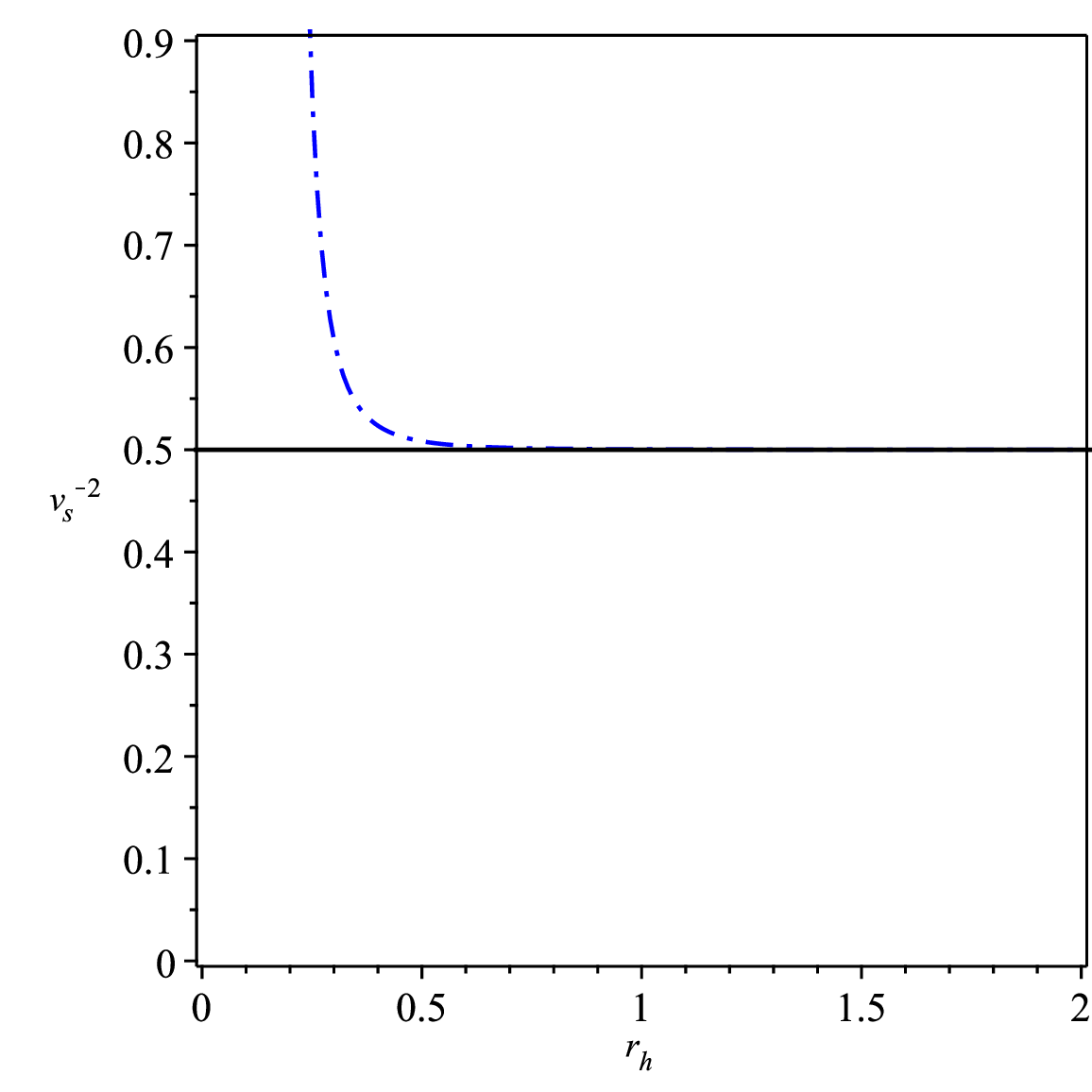}
    \caption{$ v_{s}^{-2}$ Vs.  black hole horizon with $\zeta=1$, $\alpha_{1} = \alpha_{2}=2$ and $l=2$. Here $\gamma=0$ is denoted by a black line and $\gamma=0.5$ is denoted by a blue dash-dot line.}
    \label{fig:9}
\end{figure}

The adiabatic compressibility of a black hole is defined as \cite{dolan2011compressibility}
\begin{equation}\label{eq:54}
        \beta_{S} = -\frac{1}{V_{c}} \biggl( \frac{\partial{V{c}}}{\partial{P}} \biggl)_{S}=0.
\end{equation}
Here, we find that the adiabatic compressibility is zero.
A speed of sound can be calculated for the black hole from the given formula
\begin{equation}\label{eq:55}
    v_{S}^{-2}= \biggl( \frac{\partial{\rho}}{\partial{P}} \biggl)_{S},
\end{equation}
where $\rho$ refers to the density of the black hole. This simplifies to
\begin{equation}\label{eq:56}
    v_{s}^{-2}= \frac{1152 P^{2} r_{h}^{4} (\alpha_{1} \zeta -\frac{2 \alpha_{2}}{3})^{2} \pi^{4} \Omega_{2}^{2}-1296 \zeta^{2} P r_{h}^{2} (\alpha_{1} \zeta -\frac{2 \alpha_{2}}{3}) \gamma  \pi^{2} \kappa  \Omega_{2} +729 \zeta^{4} \gamma^{2} \kappa^{2}}{2304 (P r_{h}^{2} (\alpha_{1} \zeta -\frac{2 \alpha_{2}}{3}) \pi^{2} \Omega_{2} -\frac{9 \gamma  \zeta^{2} \kappa}{16})^{2}},
\end{equation}
where value of $v_{s}^{2}$ ranges from zero to one. From Fig. \ref{fig:9}, we find that the thermal fluctuation increases the speed of sound for smaller black holes. However, for large black holes ($r_h >>1$) speed is constant and the effects of thermal fluctuation are not significant.

\section{Van der Waals black holes}\label{sec:5}
In this section, we study the behavior of charged AdS black holes in nonlinear electrodynamics as a Van der Waals fluid. The Van der Waals equation of state describes real fluids and modifies the ideal gas equation of states as
\begin{equation}\label{eq:57}
   \Bigl( P+\frac{a}{v^2} \Bigl) (v - b) = T,
\end{equation}
where $ v = \frac{V}{N} $ is the specific volume of the fluid. In the case of a black hole,  $N = A/l_{p}^{2}$ denotes the number of degrees of freedom associated with the black hole horizon. Constant represents the interaction between the molecules of a given fluid and constant b represents the nonzero size of molecules. The specific volume of black hole \cite{rajagopal2014van} is given by
\begin{equation}\label{eq:58}
    v = 6 \frac{V_{c}}{N}.
\end{equation}
This simplifies to
\begin{equation}\label{eq:59}
v = \frac{(3 \zeta^{2}-2 \alpha_{1} \zeta +\alpha_{2} ) (48 \pi^{2} P  \zeta  r_{h}^{2} \Omega_{2} \alpha_{1} -32 \pi^{2} P  r_{h}^{2} \Omega_{2} \alpha_{2} -27 \gamma  \zeta^{2} \kappa )}{36 r_{h} \zeta^{2} \pi^{2} P \Omega_{2} (\alpha_{1} \zeta -\frac{2 \alpha_{2}}{3})}.
\end{equation}
The above equations yield the pressure as 
\begin{equation}\label{eq:60}
P = -\frac{3 \gamma  \zeta^{2} \kappa  (3 \zeta^{2}-2 \alpha_{1} \zeta +\alpha_{2} )}{4 r_{h} \Omega_{2} \pi^{2} \Bigl\{ (v -4 r_{h} ) \zeta^{2}+\frac{8 \zeta  r_{h} \alpha_{1}}{3}-\frac{4 r_{h} \alpha_{2}}{3} \Bigl\}  (\alpha_{1} \zeta -\frac{2 \alpha_{2}}{3}) }.
\end{equation}
\begin{figure}[hbt]
\centering
\subfloat[]{\includegraphics[width=.5\textwidth]{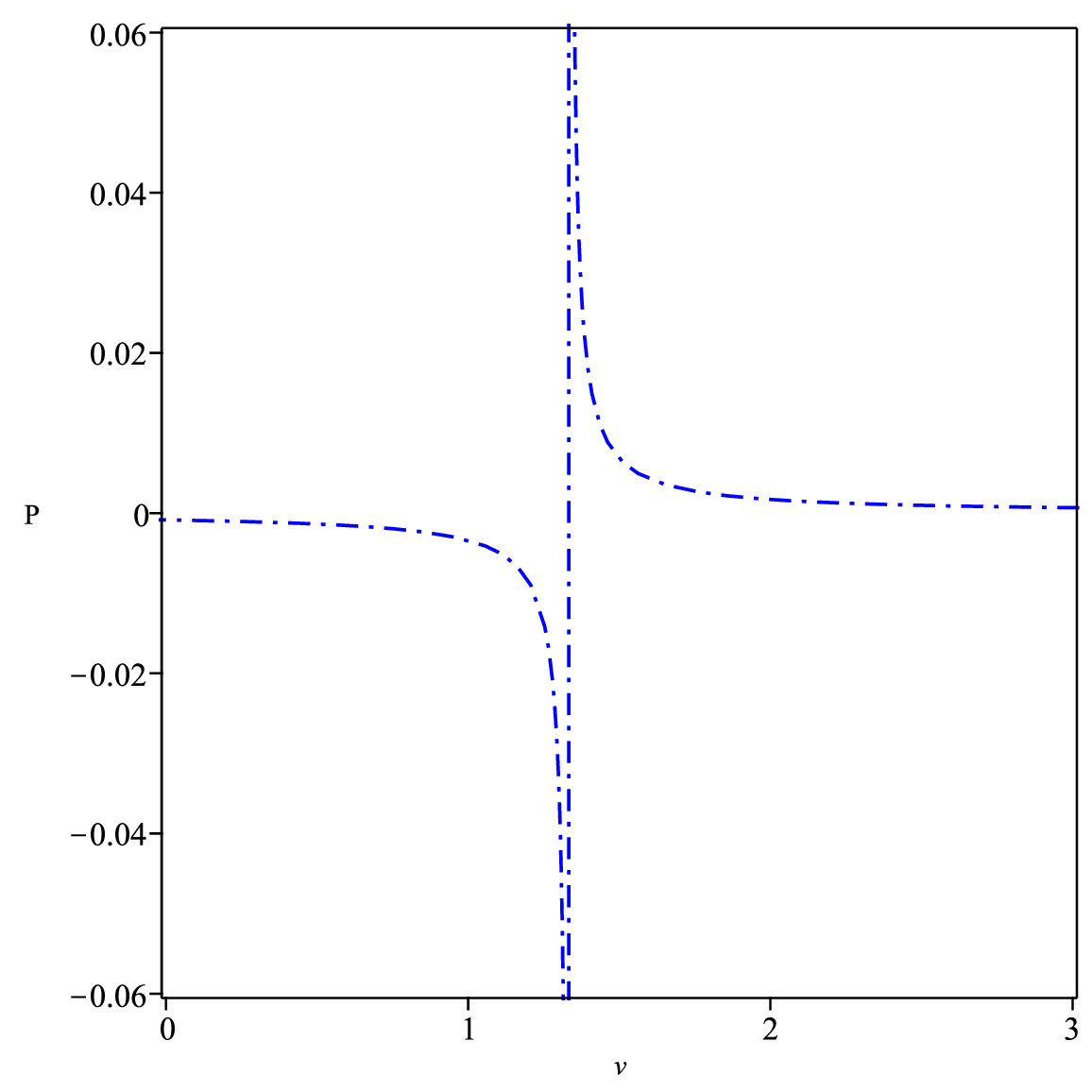}}\hfill
\subfloat[]{\includegraphics[width=.5\textwidth]{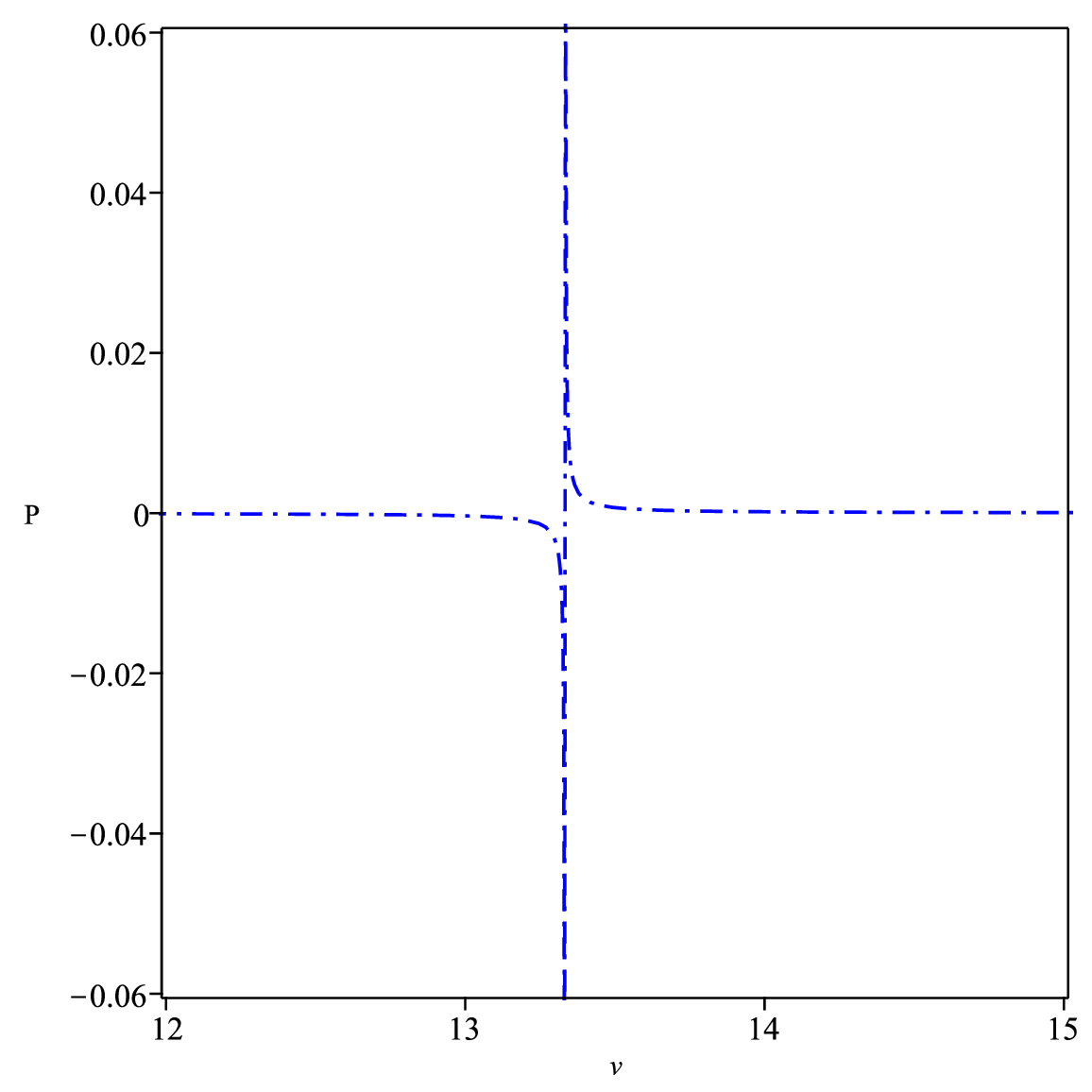}}\hfill
\caption{Pressure Vs. specific volume of  black hole with $\zeta=1$, $\alpha_{1} = \alpha_{2}=2$, $r_{h} =1$ and $l=2$.  Left panel: $r_h = 1$. Right panel:
$r_h = 10$ Here $\gamma=0.5$ is denoted by blue dash-dot line.}\label{fig:10}
\end{figure}
We plot $P-v$ diagram with $r_h = 1$ and $r_h = 10$ as depicted in Fig. \ref{fig:10}. We find that when the thermal fluctuations are not taken into account the pressure remains zero, but in the presence of thermal fluctuations a phase transition of the pressure of the black hole occurs from a negative value to a positive value. In Fig. \ref{fig:10}(a) pressure is depicted with $r_h =1$ when specific volume $v \leq 1$ pressure takes a negative value, a phase transition occurs when $v > 1$ and finally pressure takes a positive value. A similar kind of behaviour is shown in Fig. \ref{fig:10}(b) with $r_h=10$.

The equation \eqref{eq:17} leads to the pressure as
\begin{equation}\label{eq:61}
P = \frac{3 T \xi^{2}}{2 r_{h} (3 \zeta^{2}-2 \alpha_{1} \zeta +\alpha_{2} )}.
\end{equation}
Comparing equations \eqref{eq:57} and \eqref{eq:61} we can conclude that $a=0$.   Black hole mimicking the ideal gas behaviour.

\section{Conclusions}\label{sec:6}
In this paper, we have studied the effects of small statistical fluctuation on the equilibrium thermodynamics of nonlinearly charged AdS black holes in four-dimensional critical gravity. To do so, we computed the Hawking temperature for the black holes. The entropy of the black hole has an additional term at first order due to the thermal fluctuations. With the help of  Hawking temperature and corrected entropy, we have computed the 
more exact Helmholtz free energy of the black hole. The corrected Helmholtz free energy of the black hole takes a negative value. The equilibrium Gibbs free energy of the black hole is positive for small-sized black holes and for larger black holes, in contrast, it takes negative values. The corrected Gibbs free energy of the black hole is positive. We have found that
the black hole is stable in absence of thermal fluctuations. However, in presence of  thermal fluctuations,   the 
small-sized black hole becomes unstable and the large-sized black hole remains stable. Incidentally, the internal energy has not found any correction at the leading order.

On the other hand, we have also computed the corrected isothermal compressibility for the black hole. The equilibrium isothermal compressibility has found a constant negative value. However, when thermal fluctuations are taken into account, the isothermal compressibility takes a positive value for small-sized black holes and a negative for large black holes.
Finally, we have studied the $P-v$ diagram of a black hole and found that  the thermodynamic pressure vanishes for the system in equilibrium. However, the thermodynamic pressure is negative/positive for small/large-sized (with respect to specific volume) black holes with non-vanishing correction parameters.

\section*{Acknowledgements}
This research was funded by the Science Committee of the Ministry of Science and Higher Education of the Republic of Kazakhstan (Grant No. AP19674521).   D.V.S.  thanks University Grant Commission for the Start-Up Grant No. 30-600/2021(BSR)/1630.

\end{document}